# Integrating Inspection and Test Processes based on Context-Specific Assumptions

Frank Elberzhager
Frank.Elberzhager@iese.fraunhofer.de
Fraunhofer IESE, Germany

Jürgen Münch
Juergen.Muench@cs.helsinki.fi
University of Helsinki, Finland

Dieter Rombach
Dieter.Rombach@iese.fraunhofer.de
Fraunhofer IESE, Germany

Bernd Freimut
bernd.freimut@de.bosch.com
Robert Bosch GmbH, Germany

**Abstract**

Inspections and testing are two of the most commonly performed software quality assurance processes today. Typically, these processes are applied in isolation, which, however, fails to exploit the benefits of systematically combining and integrating them. In consequence, tests are not focused based on early defect detection data. Expected benefits of such process integration include higher defect detection rates or reduced quality assurance effort. Moreover, when conducting testing without any prior information regarding the system's quality, it is often unclear how to focus testing. A systematic integration of inspection and testing processes requires context-specific knowledge about the relationships between inspections and testing. This knowledge is typically not available and needs to be empirically identified and validated. Often, context-specific assumptions can be seen as a starting point for generating such knowledge. Based on the In$^2$Test approach, which uses inspection data to focus testing, we present in this article how knowledge about the relationship between inspections and testing can be gained, documented, and evolved in an analytical or empirical manner. In addition, this article gives an overview of related work and highlights future research directions.

## 1. Introduction

During the development of software, various processes are typically followed, such as different quality assurance processes. Those processes are often performed in isolation, i.e., they are applied in sequence without exploiting synergy effects between them, resulting in high effort and costs. The effort required to perform quality assurance processes, especially testing, can amount to a significant percentage of the overall development effort [1]. However, in order to be able to integrate different quality assurance processes in an effective and efficient manner, profound knowledge about the context-specific relationships between such processes is required.

We have integrated two established quality assurance processes, namely software inspections and testing [2][3]. The basic idea of the **in**tegrated **in**spection and **test**ing approach In²Test is to use inspection defect data to focus testing activities on those parts of a system that are expected to be defect-prone, or on defect types that are expected to occur during testing. Consequently, knowledge about the relationships between inspection and testing processes is necessary in order to exploit the benefits of the integrated approach, such as reduced effort or higher defect detection rates. If such knowledge is available, it can be used to control testing activities based on inspection results. However, the relationships between test processes and earlier quality assurance processes often remain unclear or are inconsistent. Therefore, assumptions have to be made. An exemplary assumption in the area of quality assurance might be that defects that are not found during inspections are also not found during white-box tests, and therefore, another testing technique should be combined with inspections [4].

A lot of different empirical models exist that present guidelines on how to gather new knowledge and derive new assumptions (e.g., [5][6][7]). In general, new relationships can be derived based on observations, which result in assumptions that are subsequently evaluated. Finally, theories may be stated that explain the evaluated assumptions in order to understand the observed relationships. However, Sjoberg et al. [7] conclude that almost no theories specific to software engineering are reported in the literature. More specifically, Jeffery and Scott [6] state that no clear theories exist with respect to software inspections. Moreover, Bertolino [8] states that for testing, no universal theory exists either. Therefore, integrating inspection and testing processes requires a sound analysis of assumptions when applying the approach for controlling testing activities.

While previous publications introduced and evaluated the applicability and initial efficiency improvements of the In²Test approach [9][10], this article concentrates on the underlying empirical concepts (e.g., definition, description, evaluation of assumptions) that were adapted for the In²Test approach from existing empirical concepts. Due to the fact that knowledge about the relationships between inspections and testing is often widely missing, a systematic approach to gathering and evaluating such knowledge is needed in order to enable people to apply the In²Test approach.

This article is structured as follows: Section 2 presents an overview of related work and analyzes the extent to which relationships are considered in existing integrated quality assurance approaches. Section 3 introduced the basic concepts of the In²Test approach before the underlying empirical concepts are refined in Section 4, which is substantiated by examples from the previously performed case study [9][10] in which the In²Test approach was validated. Finally, Section 5 concludes this article and presents future work.

## 2. Related Work

### 2.1 Integrating Static and Dynamic Quality Assurance

The main goals of the integration of static and dynamic quality assurance are reduced costs and efforts, improved defect detection and thus, improved quality, or defect prediction. Integrating such processes, i.e., using the output of one quality assurance technique as input for the next one, has gained increasing attention during the past five years [11]. A lot of different static and dynamic techniques are integrated, such as symbolic execution, testing, and runtime analysis [12], theorem proving, test case derivation and execution [13], or model checking and model-based testing [14]. For most of these approaches, only initial evaluation results exist regarding, for example, defect distributions and how well they are addressed within the approaches.

With respect to integrating inspection and testing techniques, only few approaches exist. They mainly focus on using inspections to derive test cases that can be used in later testing activities [15]. Furthermore, Harding [16] describes from a practical point of view how to use inspection data to forecast the number of remaining defects and how many defects have to be found and removed in each testing phase. One main assumption in this approach is that defects found during inspection can be used to predict remaining defect content. However, only some rules of thumb are given. Other defect prediction approaches that use inspection defect data are capture-recapture models [17], detection profile methods [18], and subjective estimations [19]. While such approaches are mainly used to decide if a re-inspection should be performed, a decision on how many tests to perform is also conceivable. However, no information is currently given on how such information can be used.

When we consider not only approaches that integrate inspection and testing techniques, but also approaches that combine such techniques in a compiled manner, i.e., the execution of inspections and testing in sequence without any exchange of information between them, additional approaches can be found. An overview is given by Elberzhager et al. [11].

A lot of studies and experiments have been performed to compare inspection and testing techniques [20], often followed by the suggestion to apply both, for example in order to maximize defect detection. Endres and Rombach [5] refer to empirical evidence that "a combination of different V&V methods outperforms any single method alone." Wood et al. [21] and Gack [22] analyzed different combinations of inspection and testing techniques and analyzed their benefit. It could be shown that in terms of cost and found defects, a mixed strategy often outperforms a strategy where only one technique is applied.

If we look only at the defect detection rates in inspection or testing, effectiveness values of between 10 to 90 percent can be found depending on the evaluation results from different environments. Consequently, it seems even more unclear what the concrete benefits are in a certain environment when a combined or integrated approach is followed. This means that context factors are often not considered systematically, and assumptions are usually not clearly stated in advance and analyzed subsequently.

## 2.2 Focusing Testing Activities

Different approaches for focusing testing activities exist, for instance the use of product metrics such as size or complexity for selecting those code classes or parts of a system that are expected to be most defect-prone [23][24]. Several studies have shown that size has an influence on defect proneness [23][25]. However, the results in those studies were not consistent, i.e., some studies showed that small code classes tend to be more defect-prone [23], and some studies showed that large code classes tend to be more defect-prone [25]. Context analyses are widely missing. Emam et al. [25] suggested not using only size when predicting defect proneness.

Another metric frequently used to predict defect-prone code classes and thus, to prioritize those parts for testing activities is McCabe complexity [26]. Moreover, Basili et al. [27] were able to show that different object-oriented complexity metrics appear to be useful to predict classes that are more defect-prone than others.

However, the application of one single metric to predict defect-prone parts of a system can lead to inaccurate predictions. Among others, Ostrand and Weyuker [23] state that "various other factors beside size might be responsible for the fault density of a particular file" and Fenton and Neil [28] suggest not using complexity as the sole predictor of defects. Consequently, a combination of metrics can lead to more accurate predictions of defect proneness. For example, Schröter et al. [29] investigated different code factors for the Eclipse programming environment, such as code complexity, process factors, and human factors, and compared them with defect-prone modules. They concluded that defect proneness can be explained by a combination of different factors; however, it is still unclear which combinations of factors, respectively metrics, are the best ones, i.e., the relationships between a combination of factors and defect proneness remains unclear.

Historical data are another source for prioritizing defect-prone parts. A number of studies have shown the relationships between different indicators, such as 'number of performed changes' or 'historical defect content', and defect proneness [30][31]. Turhan et al. [32] performed an analysis of 25 large projects in a telecommunication environment in order to focus testing based on cross-company data. However, it often remains unclear which kinds of historical data are best for prioritizing parts to be tested, i.e., relationships remain unclear or are only valid in a specific context. Finally, the experience and knowledge of experts or risk-based approaches can support prioritization of those parts of a system under test that are expected to be most critical [33].

Another kind of focusing is test case prioritization. Elbaum et al. [34] describe different techniques for prioritizing test cases in order to find a cost-efficient and effective test suite for regression testing. A number of empirical studies are summarized, new research questions are derived, and additional empirical validation is performed. The authors concluded, for example, that the effectiveness values of various techniques vary significantly across different software programs. Recent studies have confirmed this kind of focusing approach as reasonable in different environments, such as service-oriented business applications [35], and present additional knowledge about relationships between different test case prioritization techniques and defect proneness.

A third focusing concept consists of prioritizing defect types for testing. According to a defect classification used to classify defects found, testing might be focused on such defect types that appeared most often, leading to test case derivation in order to cover those defect types. For instance, the Orthogonal Defect Classification (ODC) by IBM [36] provides so-called triggers, which represent information on the environment or condition that had to exist for defects to appear. Mapping the triggers to defect types could provide guidance on where and what defect types to focus testing activities on and thus, which test cases to derive. However, a defect classification has to be chosen carefully if defect types found during different quality assurance activities are used to focus testing activities. For example, if classified inspection defects are used to prioritize certain defect types for testing, it has to be ensured that those defect types can be found during testing. With regard to the question of whether inspections and testing are complementary quality assurance activities, i.e., the question of whether they will find different kinds of defects or not, the results from experiments and case studies are not consistent. Different empirical evaluations show that inspection and testing activities are able to find defects of the same defect types [37], while others showed that inspectors and testers find different kinds of defects [38]. Some defect types might only be found either by an inspection activity or by a testing activity. However, except for obvious defect types (e.g., bad or missing comments with inspections, performance problems with testing), it remains unclear which defect types are found best with inspections and which with testing.

In general, context-specific evidence on how to focus testing is often missing or of limited validity. Furthermore, such approaches do not use data from early quality assurance activities. Especially inspection data is usually not used systematically to focus testing activities.

## 3. The In$^2$Test Approach

The In²Test approach [9][10] uses inspection data (i.e., a defect profile consisting of quantitative defect data and defect type information gathered from an inspection) to focus testing activities. The focus can be placed on certain parts of a system under test or on defect types (which is called one-stage prioritization [9]), or on both (which is called two-stage prioritization, meaning that parts of the system under test that are expected to be defect-prone are prioritized first, followed by prioritization of defect types that should be focused on in the prioritized parts [10]). The main idea of prioritization is to predict those parts for testing that are expected to be most defect-prone, or to predict those defect types that are expected to show up during testing activities.

In order to be able to focus testing activities based on an inspection defect profile, relationships between defects found in the inspection and defects to be found with testing have to be considered, which also counts for defect types. For that reason, assumptions need to be defined explicitly as such relationships are often not known. Furthermore, assumptions are often too coarse-grained to be applied. Consequently, refined selection rules have to be derived in order to be operational. For example, one assumption for the In$^2$Test approach can be a Pareto distribution of defects, i.e., parts of a system where a high number of defects are found with inspections indicate that more defects are expected to be found with testing. In this case, it has to be clarified what "high number of defects" means in a concrete environment, i.e., a concrete metric and thresholds have to be defined. Selection rules to be chosen depend on the available and analyzed data from the concrete context.

The concrete selection rules chosen also depend on explicit context factors. For example, consider the number of available inspectors and time as two context factors. If only one inspector is available who has to inspect certain parts of a system within a limited period of time, it could be hypothesized that fewer parts need to be inspected. Consequently, more effort should be spent on testing activities. Another example: Consider the experience of the inspectors as a context factor. If the inspectors' experience is low, it could be hypothesized that not many critical defects will be found. Consequently, the inspected parts should be tested again. In contrast, if the inspectors' experience is high, it could be hypothesized that most of the defects will be found before testing, and the inspected parts can be skipped for testing (in this example, it is assumed that both inspection and testing activities are able to find the same defects). Again, it depends on the concrete environment which selection rules are chosen. If a selection rule, respectively the corresponding assumption turned out to be valid, its significance increases. Both context and significance are summarized in the scope of validity.

After prioritization has been completed, test cases have to be created or selected. The number of test cases derived must fit the overall quality assurance strategy and the way prioritization is performed. For example, one strategy might be that only the top-prioritized parts of a system and defect types should be tested, omitting the remaining ones and thus saving testing effort. Another strategy might be to focus most of the available test effort on the prioritized parts of a system, and to test the remaining parts with little effort. Consequently, the aim is to improve the defect detection ability (i.e., effectiveness). Finally, the prioritized testing activity is performed accordingly.

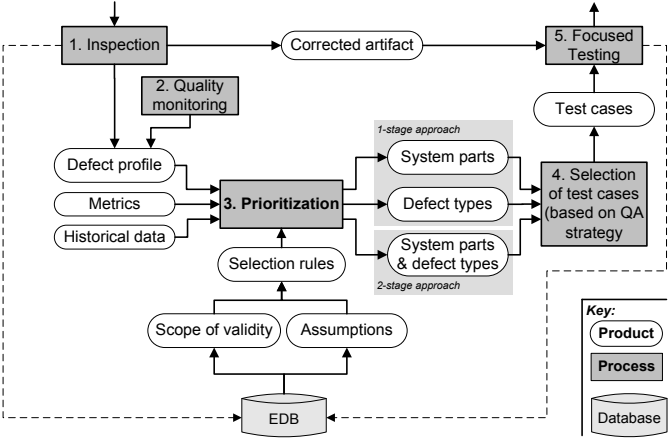

**Figure 1: The integrated inspection and testing process**

Figure 1 gives an overview of the process steps. After the inspection has been performed (step 1), a defect profile is derived, which contains, for example, the number of defects found per code class or the number of defects found per defect type. Next, quality monitoring (step 2) is done in order to ensure that the inspection data has the appropriate quality to be used for prioritization. For example, the number of defects found or the reading rate of the inspectors can be checked. One way to perform quality monitoring is to use context-specific historical data. If such data is not available, data from the literature can be used instead.

Next, prioritization (step 3) of system parts, defect types, or both is performed. For this, the defect profile is used and, optionally, metrics and historical data, which can be combined with inspection defect data in order to improve prioritization. Assumptions and refined selection rules derived from them use this data and information to perform prioritization. The assumptions defined should express the relationships in the given context between inspection defect data and the remaining expected defect distribution to be addressed with testing activities. This means that the test process is defined and directed based on the selected assumptions. If, for example, the significance of a Pareto distribution for defects is high based on past projects, this assumption can be used to focus testing activities on those parts of a system where inspection defects appeared most often.

After prioritization, the development or selection of test cases comprises step 4. Finally, the prioritized testing process is conducted (step 5) and defect results from the current testing activities can be analyzed continuously in an ideal case. If more defects are found in the selected parts, the assumptions appear to have been valid and the testing activities can continue to focus on the prioritized parts. If not, the assumptions have to be adapted and another prioritization has to be performed (i.e., a re-direction of the test process is conducted). Furthermore, parts that were not prioritized can be tested in order to check if they are defect-free, which leads to stronger empirical evidence for the assumptions.

In case no runtime evaluation is possible, at least a retrospective analysis of the assumptions should be performed. The results should be packaged in an experience database (short: EDB) and used for future prioritizations and analyses.

## 4. Empirical Concepts used the In$^2$Test Approach

This section gives an overview of how context-specific assumptions can be identified, how such assumptions can be described, and how assumptions can be evaluated. For this, existing empirical concepts are considered and adapted to the In$^2$Test approach. The In$^2$Test approach was already validated [9][10], and we re-use results from that study for demonstrating the underlying empirical concepts of the In$^2$Test approach. Thus, the following research questions (RQ) are considered:

RQ1: How can assumptions be derived and defined systematically in the context of the In$^2$Test approach?

RQ2: How can assumptions be operationalized for application in the In$^2$Test approach?

RQ3: How can the significance of the gathered evidence on the relationships between inspections and testing be increased and how can the scope be extended?

### 4.1 Identification of Context-specific Assumptions for the In$^2$Test Approach

Several models exist that guide the identification of relationships between different processes, which is usually an iterative procedure. For instance, Jeffery and Scott [6] propose a model that starts with observing and understanding a phenomenon in the real world, which is explained by a derived theory. Such a theory is evaluated, refined, and re-evaluated in order to improve the validity of the theory. Another model is given by Endres and Rombach [5]. It starts with observations in a certain environment, which can result in laws if they reappear. Those laws may predict future observations and can lead to a theory that explains the law. Finally, a theory can be confirmed, refined, or rejected based on future observations.

Various other models have been proposed in recent years (e.g., [7]). In summary, three basic steps can be distinguished independent of the concrete model: First, a certain observation is made in a given environment; second, an assumption is stated that describes certain relationships and has to be evaluated; and third, a theory is derived that explains the assumption, respectively an evaluated assumption. Figure 2 shows an overview of these steps. Due to frequent uncertainty regarding relationships, assumptions are often stated first (which might be a first guess, an experience, or lessons learned from different domains). An assumption describes context-specific relationships that are observed or appear to be useful, but are not empirically grounded. In contrast, evaluated assumptions are based on empirically valid results that are accepted in the given context. Evaluated assumptions might lead to a theory that explains them. Steps one to three have to be repeated for each new context. However, assumptions and theories might be valid in several contexts, which will have to be evaluated.

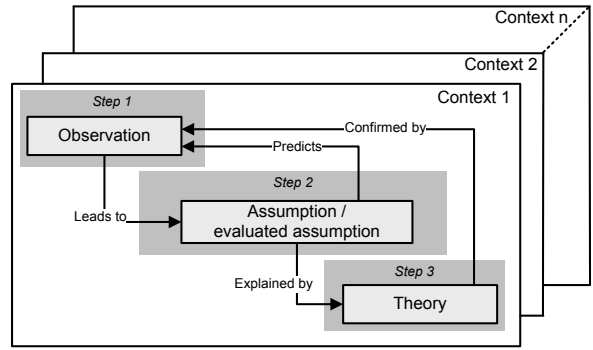

Figure 2: Observations to assumptions, from evaluated assumptions to a theory

To answer RQ1, we propose deriving assumptions in two different ways: analytically or empirically.

1. Analytically: Based on a systematic analysis of a certain environment, which includes considering process and product structures (e.g., development and quality assurance processes; experience of developers, inspectors, and testers; size and complexity of product to be developed), assumptions regarding the relationships can be derived in a rational manner. For example, consider that only parts of a system were inspected due to an insufficient amount of time available for inspections; consequently, testing should especially be focused on those parts of the systems that were not inspected, based on the assumption that inspections already find an acceptable number of relevant defects.

2. Empirically: Based on (i) empirical knowledge from different environments and (ii) new experiences from a given context, assumptions regarding relationships can be derived empirically. First, accepted empirical knowledge from different contexts can be used and – if possible – adapted to a given context. Examples in the area of quality assurance that are often found in the literature are that developers are unsuited to test their own code or that most of the defects are found in a small number of modules. Such empirically proven knowledge can be adapted and a corresponding assumption has to be checked in the given context. Second, by studying certain processes in a given context, new observations may be made, resulting in new or refined assumptions. This means that new empirical knowledge about certain relationships is gained. For example, when classifying defects according to a certain defect classification, new insights about which defect types are found by different quality assurance activities might be obtained.

### *4.2 Structured Description of Assumptions for the In$^2$Test Approach*

To answer RQ2, this section presents a structure for describing experiences made with quality assurance processes. In addition, it describes how assumptions can be made operational. This is exemplarily shown by embedding them into so-called selection rules, which are used for integrating inspections and testing.

In order to be able to understand various processes, and consequently improve them, knowledge regarding the relationships between such processes is required. Such knowledge about relationships can be seen as experience. In general, experiences are valid within a certain scope. This scope can be characterized by the context (i.e., the environment with all kinds of factors that might have an influence on the experience) and the degree of validity of the experience (simplified, this means how many times the experience has been gained or confirmed). In order to emphasize that there are many different experience items, we call one item an experience element [39].

In the area of inspection and testing, for instance, an exemplary assumption might be the Pareto distribution mentioned above (context and significance are initially not considered here):

*Assumption 1: Parts of a system where a large number of inspection defects are found indicate more defects to be found with testing.*

The assumption was evaluated in a case study [9][10] and evidence was gained for its validity in that context. Selection rules operationalize assumptions so that they can be applied. Different concrete selection rules that make the assumption operational are conceivable, such as:

*Selection rule 1: Focus a testing activity on those system parts where the inspection found more than 20 defects.*

*Selection rule 2: Focus a testing activity on those system parts where the inspection found more than 8 major defects.*

*Selection rule 3: Focus a testing activity on those system parts where the inspection found more than 15 defects per 1000 lines of code.*

*Selection rule 4: Focus a testing activity on those system parts where the inspection found more than 12 defects and which contain more than 600 lines of code.*

*Selection rule 5: Focus a testing activity on those system parts where the inspection found more than 12 defects and which contained more than 20 defects in the two past releases.*

A selection rule consists of an action that describes what to do, and a precondition that describes what has to be true in order to perform the action. For example, an action may be to focus a testing activity such as unit testing on certain code classes. Furthermore, a precondition consists of a logical expression and a concrete threshold, and uses inspection defect data (e.g., selection rules 1-3), and optionally metrics (e.g., selection rule 4) or historical data (e.g., selection rule 5). Such thresholds must be defined in a context-specific way and can only be valid for that context. One way to define a concrete threshold is to take the part where the inspection found most defects and use the 80% value of this number.

With respect to the validity of assumptions and derived selection rules, each one has to be evaluated in a new environment in order to identify the most suitable ones in a given context, i.e., the scope of validity has to be determined.

Figure 3 summarizes the concepts of how experiences can be described in a structured manner using selection rules that cover knowledge expressed in the form of assumptions, respectively evaluated assumptions.

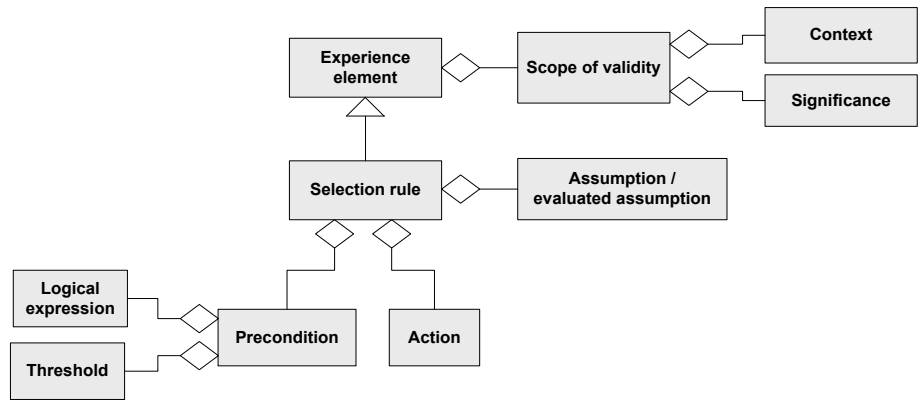

**Figure 3: Structural model of relationships**

*4.3 Evaluation of Selection Rules and Assumptions for the In²Test Approach*

This section provides an answer to RQ3 by describing how assumptions that are initially stated should be refined, i.e., how they should be evaluated by gaining new empirical evidence and considering additional context factors.

### *4.3.1 Context check*

In order to achieve a significant reduction in effort for a prioritized testing activity with appropriate quality, the assumptions stated should be evaluated during inspection and testing activities. This is especially relevant with respect to context factors that might change during the development and quality assurance activities.

Consider the following example. The assumption is made that code classes contain additional defects in which the inspectors have already found a significant number of defects. Consequently, these code classes should be prioritized for testing. One underlying context factor is the inspectors' experience. In the exemplary environment, the assumption is only true when inspectors with little experience conduct the inspection. Thus, when a project manager or quality assurance engineer plans testing activities, he has to consider the context factors. If, for example, only inspectors with little experience are available, the assumption may be true and he can prioritize those code classes for testing in which the inspectors have found defects. However, it may happen that suddenly inspectors are available with a lot of experience, and the assumption made initially in this case is not true, because in past quality assurance runs, it appeared that experienced inspectors find most of the problems in the inspected code classes and testing does not reveal many new defects. Consequently, if such a change of context factors occurs and if the project manager, respectively quality assurance engineer, knows about the influence of a context factor on the relationships, he is able to adapt the testing strategy with respect to the changed context factors.

### *4.3.2 Maintenance of evidence*

In order to be able to decide which assumptions and derived selection rules are suitable in a given context in order to allow focusing testing activities based on inspection results, a retrospective analysis is necessary. For conducting such an analysis, data gathered during the quality assurance run have to be considered. This comprises at least inspection defect data and test defect data. For a detailed analysis, a representation of the number of defects found per part has to be given, e.g., the number of defects found per code class. Furthermore, if the defect type is considered in the assumptions, the number of defects found per defect type is needed. Finally, if additional metrics are used (e.g., size, complexity), these data must also be captured, e.g., size – measured in lines of code – per code class.

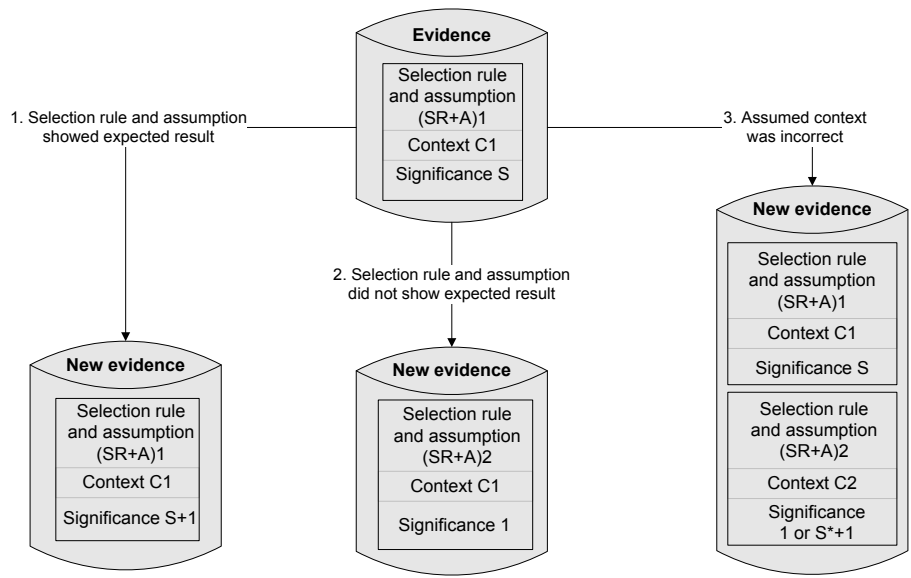

**Figure 4: Maintenance of evidence (based on [40])**

Figure 4 gives an overview of the three possibilities when selection rules and assumptions are evaluated in order to assess how they evolve. An exemplary model for packaging project experience [40] is used to store a set of selection rules and assumptions, respectively their performance, in the completed quality assurance run. The starting point for focusing testing activities is the choice of selection rules and assumptions (e.g., from a database) before they are applied in a new project. The selection rule and the assumption (SR+A)1 used are valid in a given context C1 and have a certain significance gained from application in S former projects. From our experience, only a small set of context factors has a considerable impact on the validity of selection rules, and usually experienced practitioners from a certain context know such context factors quite well. Of course, if a selection rule is applied the first time, its significance is zero. An analysis with respect to the gathered data can lead to three different possibilities:

1. The selection rule and the corresponding assumption were correct during the completed quality assurance run, i.e., focusing testing activities based on inspection results was appropriate and led to the expected results (i.e., all defects were found). In this case, significance is increased by one.

2. The selection rule and the corresponding assumption were incorrect during the completed quality assurance run, i.e., focusing testing activities based on inspection results was not appropriate and did not lead to the expected results (i.e., defects were not found). In this case, an alternative assumption and an alternative selection rule (SR+A)2 must replace the original ones used in the given context. Significance is set to one.

3. The project context was different for the completed quality assurance run, i.e., concrete values of certain context factors were assumed (e.g., low experience of inspectors), but after following the processes, this turned out to be wrong due to hidden or changed context (e.g., the experience of the inspectors was actually high). Consequently, the original selection rule and assumption are kept as is, and a new selection rule and assumption (SR+A)2 are used in the changed context with significance 1 or S*+1 if that selection rule was already applied successfully in previous quality assurance runs.

Major conclusions from such an analysis should be taken into account in subsequent quality assurance runs.

In order to perform such a maintenance analysis of the assumptions, and taking into account lessons learned from analyzing the assumptions and selection rules during the initial case study [9], the maintenance model was adapted for the In$^2$Test approach. In order to be able to judge the quality of selection rules and to decide which assumptions are most appropriate, a four-scale evaluation scheme is introduced to assess the selection rules. We only consider the first two cases of the model explicitly here (i.e., assumptions and selection rules showed the expected results, respectively did not show the expected results) and omit the third case (i.e., the assumed context was incorrect), as the third case is very similar to the second one (except for the new context). The main reason for defining such a classification is that it allows judging each rule with respect to its effectiveness and efficiency, and consequently permits getting initial feedback about the suitability of the rules.

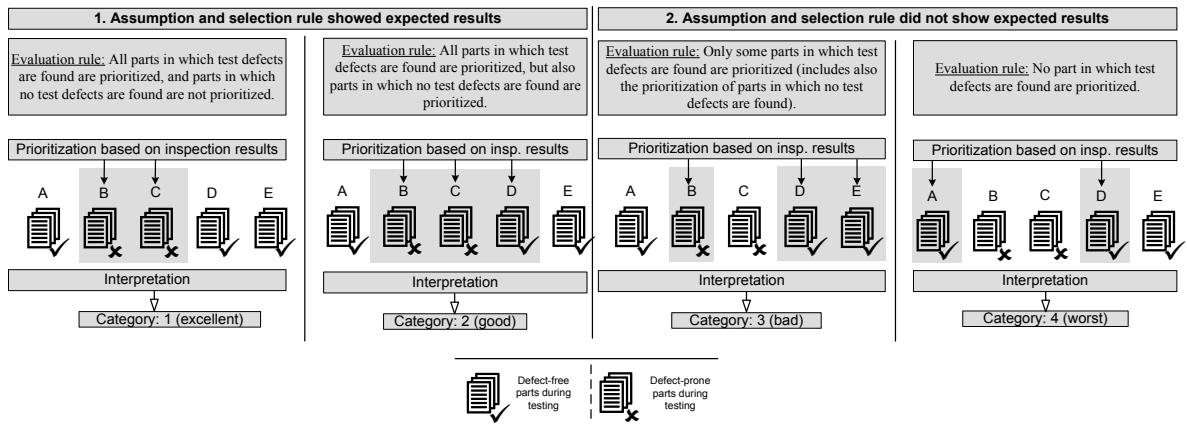

Figure 5: Four quality categories

If an assumption and the derived selection rule showed the expected results, two refined possibilities for their evaluation exist, i.e., a selection rule can be classified in category one or two if all defects are found with prioritization (i.e., the selection rule is correct). The two categories differ with regard to the selection of additional parts (e.g., code classes), which would result in lower efficiency. If an assumption and the derived selection rule did not show the expected results, there are also two possible evaluations. Categories three and four comprise selection rules that select none or only some of the defect-prone parts (e.g., code classes), which results in reduced overall effectiveness (i.e., the selection rule is incorrect, which means that not all or none of the defects are found with the chosen selection rule). Figure 5 shows examples of each of the four categories.

If the In$^2$Test approach is applied to reduce the effort for testing activities, still no defects should remain uncovered (i.e., strong evaluation rule), respectively a certain threshold should not be exceeded (i.e., weak evaluation rule) when applying a selection rule that selects only a subset of all possible parts under test. Thus, selection rules classified into one of the first two categories represent selection rules of the highest quality. A selection rule that prioritizes all system parts (e.g., code classes) would also have been placed into category two. However, this case is excluded because no effort reduction would be achievable. The third category overlooks defect-prone parts, but selects at least some defect-prone parts. Thus, a combination or consideration of selection rules of this category might improve the prioritization of defect-prone parts and, consequently, should be further analyzed in future quality assurance runs. Finally, selection rules of category four do not lead to any appropriate prioritization and are thus negligible.

Besides the four-scale evaluation scheme, which provides an initial overview of the quality of the selection rules, there are other ways of judging the quality that also make sense. If stronger differentiation between rules regarding their quality should be provided, precision and recall can be calculated for each rule, and an aggregated F-measure can be calculated

[42]. In this context, precision expresses the quality of a prioritization, and recall its completeness. If such values are calculated, those selection rules that lead to the highest overall F-measure are promising candidates for future prioritizations.

Table 1: Three assumptions, refined selection rules, and quality categories based on case study results [9]

| Assumptions and selection rules | | Selection of code classes | Quality category | Precision | Recall | F-measure |
|---|---|---|---|---|---|---|
| *Assumption 1: Parts of the code where a large number of inspection defects are found indicate more defects to be found with testing.* | | | | | | |
| No. | Selection rule: Focus testing on those code classes in which the inspection determined.. | | | | | |
| A1.01 | defect content > 25 | II, III | 2 | 0.5 | 1 | 0.67 |
| A1.02 | defect density > 0.05 | I, III, IV | 2 | 0.33 | 1 | 0.5 |
| *Assumption 2: Parts of the code where a large number of inspection defects are found and which are of large size indicate more defects to be found with testing.* | | | | | | |
| No. | Selection rule: Focus testing on those code classes in which the inspection determined.. | | | | | |
| A2.01 | defect density > 0.05 & class length > 500 LoC | III | 1 | 1 | 1 | 1 |
| *Assumption 3: Parts of the code where a large number of inspection defects are found and which are of small size indicate more defects to be found with testing.* | | | | | | |
| No. | Selection rule: Focus testing on those code classes in which the inspection determined.. | | | | | |
| A3.01 | crash severity (defect content > 0) & mean method length < 10 | I, IV | 4 | 0 | 0 | 0 |

Table 1 shows an excerpt of an evaluation based on the case study results given in [9], and presents both classification schemes. Three exemplary assumptions and refined selection rules are stated, which are evaluated based on the defect data and further metrics as stated in Table 2. For example, assumption 1 presents the already mentioned Pareto distribution, and based on the concrete threshold (which is defined context-specifically) in selection rule A1.01, code classes two and three are selected. During testing, six additional defects were found in code class three; consequently, selection rule A1.01 is categorized as '2' because the defect-prone code class has been prioritized, but code class two was selected as well, which turned out not to contain additional defects during testing. Accordingly, precision, recall, and F-measure are calculated for each rule. The results demonstrate that the best rule is A2.01 (the best possible value '1' is calculated), the worst rule is A3.01 (the worst possible value '0'), and fairly good values are obtained for A1.01 and A1.02. Regarding the latter, it can be observed that these two selection rules belong to the same quality category, but differ in terms of quality, which is given by the refined calculations.

Table 2: Defect data and further metrics from case study [9]

| | Code classes | | | |
|---|---|---|---|---|
| *Metrics* | I | II | III | IV |
| Inspection defect data | 14 | 40 | 39 | 7 |
| Inspection defect data (crash) | 1 | 0 | 0 | 1 |
| Inspection defect density | 0.06 | 0.03 | 0.06 | 0.06 |
| Test defect data | 0 | 0 | 6 | 0 |
| Class length (LoC) | 231 | 1364 | 701 | 115 |
| Mean method length (LoC) | 3.28 | 13.54 | 8.11 | 7 |

## 5. Summary and Outlook

Any integration of different software quality assurance processes requires knowledge about the relationships between these processes in order for the integration to be highly effective and efficient. This article proposed the integrated inspection and testing approach In$^2$Test and emphasized the importance of gathering empirical knowledge in this area. Sjoberg et al. [7] and Bertolino [8] stated for software inspections and testing that no clear theories exist that explain such relationships. Due to the lack of reliable theories, it is essential to substantiate research that integrates inspection and testing processes by systematically defining, describing, and evaluating assumptions, and by considering the scope of validity. With the presented mechanisms, context-specific assumptions can be defined analytically and/or empirically (RQ1). Selection rules as defined in this article refine assumptions in order to make them operational. They are intended to be suitable for a certain context with a certain degree of significance, and consist of preconditions and actions (RQ2). The presented maintenance model for evaluating selection rules and assumptions can be used to improve their validity in a given context (RQ3). The proposed concepts can be applied as long as inspection and test defect data is available (and possibly additional product metrics).

Exemplary assumptions and selection rules presented in this article can serve as a starting point for using the In²Test approach in an industrial environment. However, identifying the assumptions and analyzing their validity must be done in a sound way and might take some effort.

With respect to future research directions, more empirical evidence needs to be gathered from different environments when applying the integrated inspection and testing approach. Preliminary results from two case studies have substantiated the applicability of the In²Test approach and resulted in additional assumptions [9][10][41]. However, in order to be able to derive theories that explain relationships between inspections and testing, additional empirical evidence has to be gathered. Second, additional quality assurance activities may be included in the In²Test approach, which might, for example, lead to the use of results from static analysis tools to allow more comprehensive focusing. Third, the structure of assumptions, their derivation, and their evaluation could be used in several different environments in order to systematically cover, compare, and analyze knowledge between processes and contexts.

## Acknowledgements

This work has been funded by the Stiftung Rheinland-Pfalz für Innovation project "Qualitäts-KIT" (grant: 925). We would like to thank Sonnhild Namingha for proofreading.